\documentclass[twocolumn,preprintnumbers,amsmath,amssymb,prl,aps]{revtex4}

\usepackage{graphicx}
\usepackage{dcolumn}
\usepackage{bm}
\usepackage{epstopdf}
\usepackage{multirow}
\usepackage{color}
\usepackage{amssymb}
\usepackage{float}

\begin{document}

\title{Magnetic structure map for face-centered tetragonal iron: appearance of a new collinear spin structure} 
\author{D. Reith}
\email{david.reith@univie.ac.at}
\author{R. Podloucky}
\affiliation{
Institute of Physical Chemistry, University of Vienna and Center for
Computational Materials Science, Sensengasse 8, A-1090 Vienna, Austria
}
\author{M. Marsman}
\affiliation{
Computational Materials Physics, University of Vienna and Center for
Computational Materials Science, Sensengasse 8, A-1090 Vienna, Austria
}
\author{P. O. Bedolla-Velazquez}
\author{P. Mohn}
\affiliation{
Department of Applied Physics, Vienna University of Technology and Center for
Computational Materials Science, Makartvilla, Gu{\ss}hausstra{\ss}e 25-25a, A-1040 Vienna, Austria
}
\date{\today}

\begin{abstract}
For fcc and tetragonal distorted fct iron 
a large number of magnetic configurations as a function of crystal structural
parameters were studied by means of density functional theory concepts. The stability
of magnetic structures was defined by the magnetic re-orientation energy
$\Delta {E}^i_\text{reor}$ as the difference of the total energy of
configuration $i$ and that of the fcc ferromagnetic state.
The Cluster Expansion technique was applied to six volumes
deriving $\Delta {E}_\text{reor}$ for more than
90.000 collinear spin structures at each volume.
Structures with low $\Delta {E}_\text{reor}$ were tetragonally distorted according to a two-dimensional mesh
defined by volume per atom $V$ and $c/a$ ratio.
At each mesh point $\Delta E_\text{reor}$ 
for all collinear structures were compared to results for spin spirals (SS) which were
calculated on a grid of propagation directions, and then the lowest
$\Delta {E}_\text{reor}$ defined the magnetic structure map.
Three local minima were identified and for each of the minima SS were
calculated on a fine grid of propagation vectors. 
At the minimum with $V=10.6$\,\AA\,$^3$ and $0.94 \le c/a \le 1.01$ 
a hitherto unknown simple collinear spin structure
with four atoms per fct unit cell was the most stable one. It
consists of two atoms with
anti-ferromagnetically ordered local moments of $\pm 1.8 \mu_\text{B}$
and of two atoms with zero local moment. 

\end{abstract}
\maketitle

The intriguing magnetic orderings of fcc-related phases of Fe arrested
particular attention.  Although a large number of experimental as well
as theoretical studies were performed it is still an open question, if 
unknown phases exist.  Indeed 
we detected a 
new simple collinear magnetic ordering, which 
includes  atoms with zero local moment.

We search for new structures by means of a
map 
describing magnetic ordering versus
volume per atom $V$ and 
$c/a$ ratio of 
tetragonally distorted 
fcc Fe. 
Thereby scanning 
a large configuration space for magnetic orderings 
for which we developed a new strategy based on spin dependent total energies
as derived by 
density functional theory (DFT) calculations.  
We will address the 
fcc-related phases 
as 
fct Fe as tetragonal distortion is very important
for the stabilization of magnetic structures. 

Experiments were done on thin-films
\cite{Pescia1987,Darici1987,Liu1988,Macedo1988,Stampanoni1989,Lu1989,Landskron1991,Wutting1993}
and precipitates \cite{Tsunoda1989,Tsunoda1993,Tsunoda2007}. 
The fct structure was enforced by a host material 
or substrate with fcc structure (e.g. Cu).
Diffraction measurements on
precipitates\cite{Tsunoda1989,Tsunoda1993} observed a helical spin spiral (SS)
which stimulated
DFT studies.
\cite{Knoepfle2000,Spisak2000,Spisak2002,Sjostedt2002,Marsman2002,Abrikosov2007}
Marsman {\em et al.}\cite{Marsman2002} found the experimentally claimed
SS when the fcc structure 
was tetragonally distorted.
This was 
confirmed by a recent experiment
on precipitates.\cite{Tsunoda2007} 
Low-energy
electron diffraction (LEED) on thin-films was inconclusive
suggesting a range of hitherto unresolved magnetic configurations.
\cite{Pescia1987,Liu1988,Macedo1988,Stampanoni1989,Wutting1993}

\begin{figure}
\includegraphics[width = 0.47\textwidth]{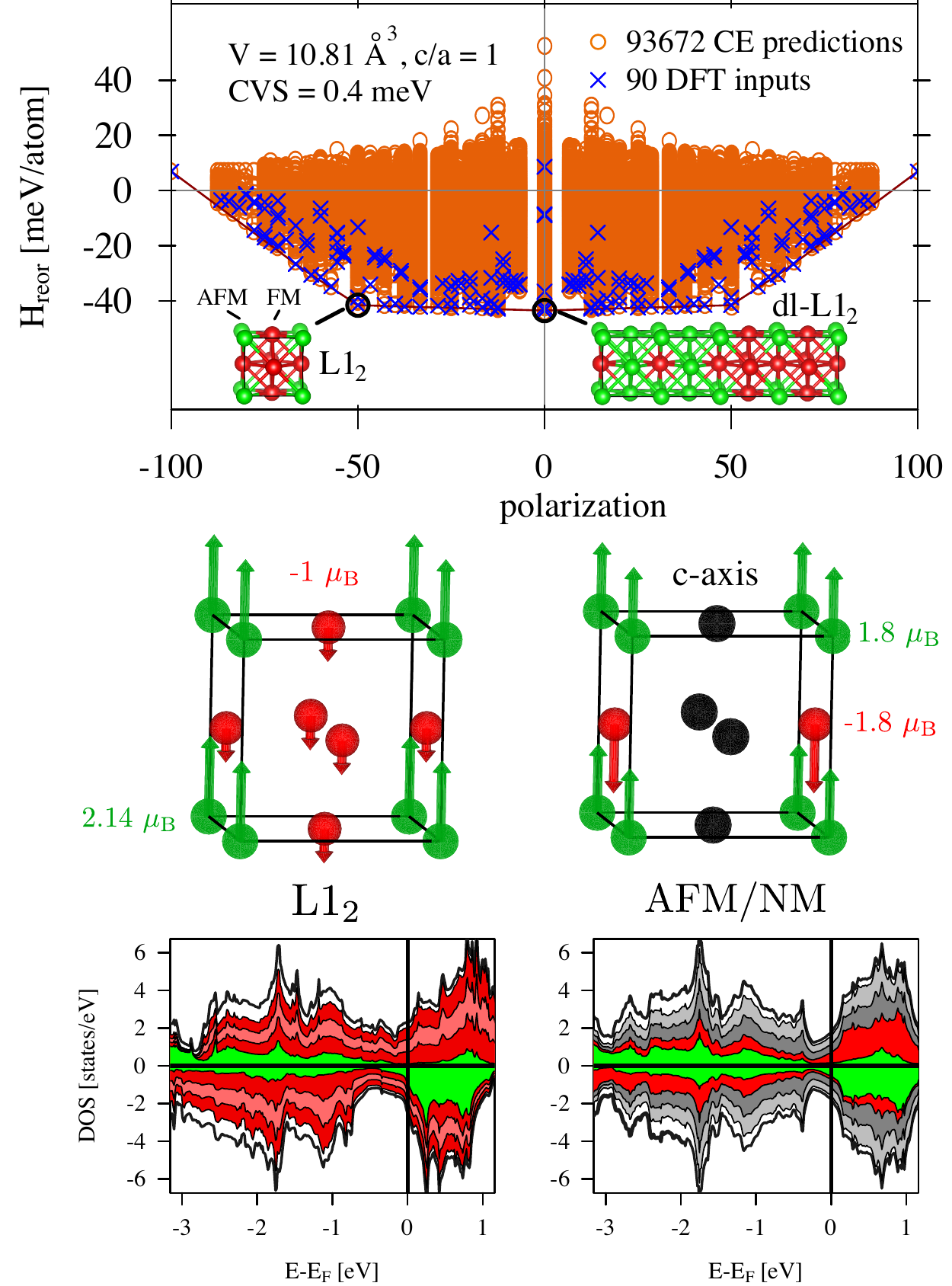}
\caption{(Color online)  
Upper panel:
CE for collinear magnetic structures 
on a parent fcc lattice for up to 16 atoms per unit cell. 
Magnetic re-orientation energy vs. total spin polarization, 
$\Delta E_\text{reor} = E(V) - E_\text{FM}(V_0)$ as defined by the total energies $E(V)$ of each spin
configuration with respect to the total energy $E_\text{FM}(V_0)$ of fcc
ferromagnetic Fe at equilibrium volume $V_0=10.52 \,\mathrm{\AA}$. 
Center panel:
Collinear L1$_2$-like magnetic configuration (left) with a cubic lattice.
It is unstable under tetragonal distortion by which the AFM/NM configuration
(right) for $c/a<1$ is stabilized (see Fig.~\ref{fig3}).
Black spheres: atoms with zero local moment.
Lowest panel: spin polarized local and total density of states (DOS). 
Local DOSes are added up subsequently.
Positive/negative values: DOS of majority/minority spin states.
}
\label{fig1}
\end{figure}

First, a large set of collinear magnetic
configurations with fcc lattice was sifted through
according to Fig.~\ref{fig1}. 
This was done by Cluster Expansion\cite{Sanchez,Ferreira,Muller,Lerch} (CE)
at several $V$s.  Then, structures were selected and
\textit{tetragonally} distorted according to a two-dimensional mesh 
defined by  $V$ and $c/a$ (see Fig.~\ref{fig4}). 
For each mesh point the magnetic re-orientation energy $\Delta E_\text{reor}$
(see caption of Fig. \ref{fig1}) for selected collinear
structures were compared to $\Delta E_\text{reor}$ for SSs with
a selected set of propagations $\vec{q}$ 
and by that three local minima were found. For each minimum SSs were re-calculated on a fine grid of
propagations. Finally, the magnetic structure with the lowest $\Delta E_\text{reor}$ is
indicated on the two dimensional structure map.
Refs.  \onlinecite{Singer2011,Dietermann2012,Faehnle2013} presented
a general  multi-spin-configuration CE, which includes
also SSs.  Such a general concept 
however is expected to be computationally hardly feasible and no applications to realistic cases have been
published until now. Furthermore, it is unclear how $c/a$ distortions 
might be included, which however, are important.

DFT calculations for spin-dependent total
energies were done by VASP\cite{Kresse1996,Kresse1999} within 
the projector augmented wave  method.\cite{Blochl:1994p8789}
The generalized gradient parametrization of
Ref.\onlinecite{PBE1996} was chosen and the basis set size  cutoff was 400 eV.  
The Brillouin zone integration was made by a Gaussian smearing technique
and the broadening of $\sigma=0.2$~eV on a $17\times 17\times 17$ Monkhorst and Pack
\cite{Monkhorst1976} $\vec{k}$-point mesh for a one atom unit cell.
For larger cells the mesh was scaled down accordingly.  
SSs were calculated by means of the generalized Bloch
theorem.\cite{Kuebler,Hobbs2000,Marsman2002} The local magnetic spin
moments were determined for a sphere of radius $R_{loc}= 1.164$ \AA\,.

Utilizing the package UNCLE.\cite{Lerch} 
a binary CE \cite{Sanchez,Ferreira,Muller,Lerch} was performed for collinear 
up/down  spin ordering  
on an fcc parent lattice at the six volumes per atom, 
$V= 10.27, 10.81, 11.18, 11.42, 11.76, 12.01$\,\AA$^3$
while
atomic positions and cell shape were not relaxed.
The condition for accepting 
a given spin configuration for the CE was that
the local moments were $\mu \geq  |0.1| \mu_\text{B}$.  The
CE fitting was done by  least-square
minimization\cite{Lu91} checking its quality in terms of
the (leave one out) cross validation score (CVS) \cite{Walle2002}. A genetic
algorithm was applied for the selection of clusters up to six-body
interactions. Due to spin interchangeability 
$\Delta E_\text{reor}$ is symmetric with respect to the total
spin polarization (see Fig.~\ref{fig1}). 

Discussing the CE calculations we focus only on 
$V=10.81$\,\AA$^3$.  The
total number of DFT input structures was 90 and  the configuration search
was done for up to 16 atoms per unit cell, resulting in 
93672 magnetic configurations. 

The CE derived ground states strongly depend on volume. At larger
volumes $V > 11.1$\,\AA$^3$ the most favorable ordering is the 
double-layer anti-ferromagnetic (dl-AFM)
configuration,\cite{Spisak2000,Spisak2002,Marsman2002,Abrikosov2007} which is
unstable under tetragonal distortion and monoclinic
shearing.\cite{Spisak2002,Marsman2002} Its stability
in comparison to SSs is disputed.\cite{Sjostedt2002,Abrikosov2007}
Remarkably, at the  smaller volume of
$V=10.81$\,\AA$^3$ a ferrimagnetic configuration resembling the
crystallographic L1$_2$ (Cu$_3$Au) structure 
in combination with  a very similar double-layer L1$_2$-like (dl-L1$_2$) is
found to be stable.
As sketched in Fig.~\ref{fig1} L1$_2$ consists of
magnetic moments of distinctly different sizes: a large moment with
$\mu=+2.14\,\mu_\text{B}$ and three small moments $\mu=-1\,\mu_\text{B}$,
resulting in the total moment of $\mu_{tot}=-0.86\,\mu_\text{B}$ per unit
cell.

\begin{figure}
\includegraphics[width = 0.47\textwidth ]{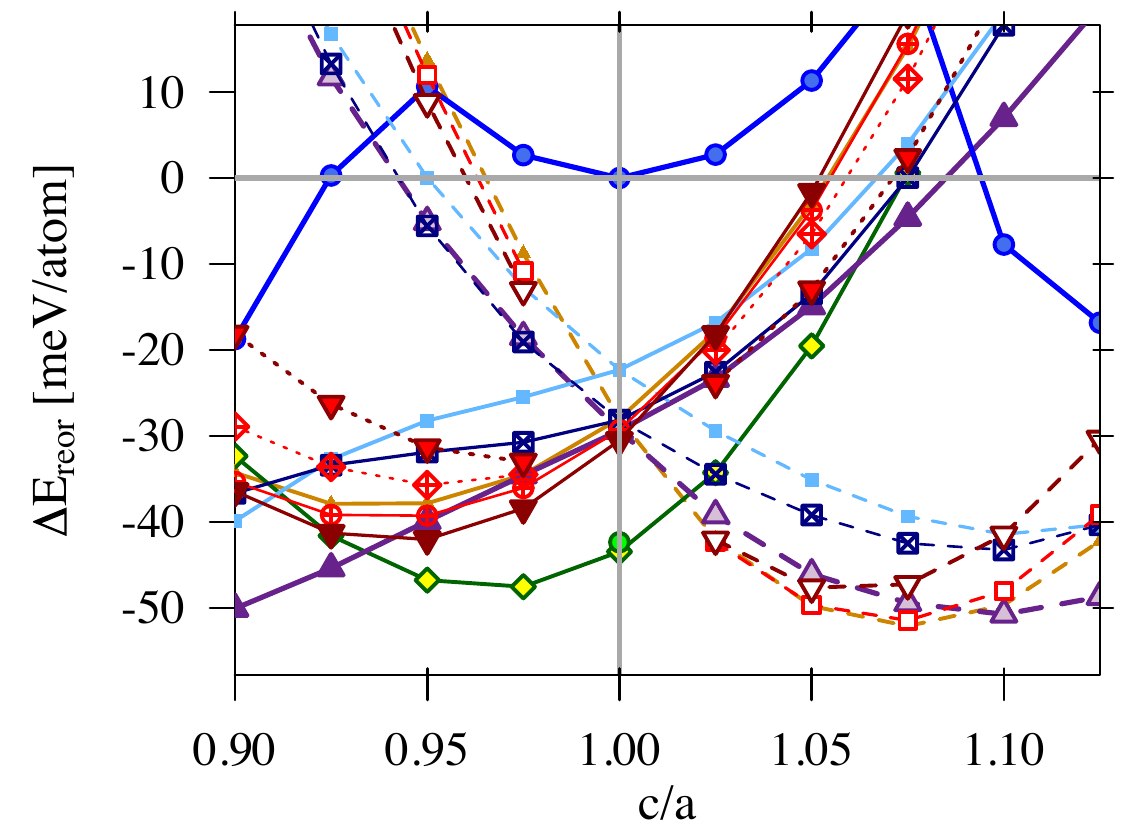} 
\caption{(Color online) 
Magnetic re-orientation energy $\Delta E_\text{reor}$ as a function of $c/a$ for a
variety of collinear configurations and SSs. For each point, i.e.
fixed $c/a$, $\Delta E_\text{reor}$ is minimized with respect to volume $V$.
Symbols refer to magnetic configurations as defined in Fig.~\ref{fig4}.
}
\label{fig3} 
\end{figure}

\begin{figure}
\includegraphics[width = 0.47\textwidth ]{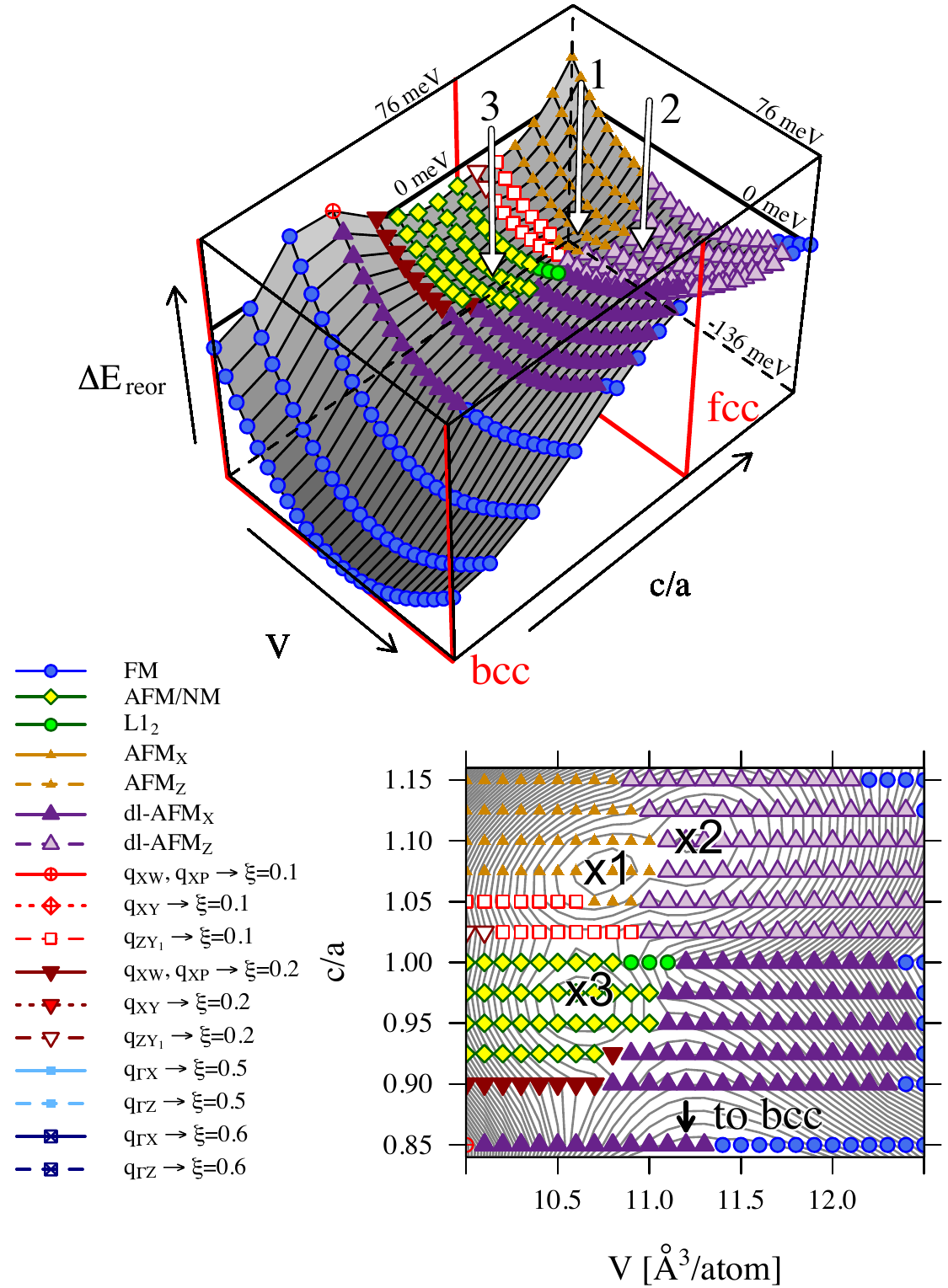} 
\caption{(Color online) 
Symbols denote the investigated collinear 
orderings
and SSs (see text).
Collinear AFM spin structures with tetragonal distortion applied  
perpendicular to the sequence of alternating spin up and down FM  planes are:  
AFM$_X$, dl-AFM$_X$; parallel to FM planes: AFM$_Z$, dl-AFM$_Z$.
Upper panel:  three-dimensional magnetic structure map defined by the lowest
$\Delta E_\text{reor}$ as a function of $c/a$ and volume $V$. 
Lower panel: two-dimensional representation of upper
panel.  Contour lines drawn in steps of 2.38 meV/atom. Local minima are
denoted by \textbf{x1}, \textbf{x2} and \textbf{x3} (see text).
}
\label{fig4} 
\end{figure}

By tetragonal distortion the  moments of 
dl-AFM ordering  and other studied anti-ferromagnetic (AFM) configurations 
remain rather unchanged.  However, for 
L1$_2$ the low moments in the ferromagnetic plane collapse
resulting in a peculiar mixed anti-ferromagnetic/ nonmagnetic (AFM/NM) spin configuration, in which
AFM planes with  moments of $\mu=1.8\,\mu_\text{B}$  alternate
with NM planes (see Fig.\ref{fig1}). 
Remarkably, even for  $c/a =1$ and
$V < 10.9$ \AA$^3$ the AFM/NM configuration is more stable by one meV/atom 
than cubic L$1_2$ (see Fig.  \ref{fig3}). 

Focusing on SSs, of interest are spirals
with propagations  $\vec{q}$ in direction $\Gamma - X/Z$, and spirals 
in direction $X/Z - P/Y/Y_1$. 
The following  propagations were considered:
$\vec{q}_{\Gamma \text{X}}(\xi)=\frac{2\pi}{a}\left(\xi,0,0\cdot({c/a})^{-1}\right)$, 
$\vec{q}_{\Gamma \text{Z}}(\xi)=\frac{2\pi}{a}\left(0,0,\xi\cdot({c/a})^{-1}\right)$,
$\vec{q}_{\text{XP}}(\xi)=\frac{2\pi}{a}\left(1,0,\xi\cdot({c/a})^{-1}\right)$, 
$\vec{q}_{\text{XY}}(\xi)=\frac{2\pi}{a}\left(1,\xi,0\cdot({c/a})^{-1}\right)$, and
$\vec{q}_{\text{ZY}_{1}}(\xi)=\frac{2\pi}{a}\left(\xi,0,1\cdot({c/a})^{-1}\right)$ 
whereby  $a$ defines the lattice parameter and $c/a$ the tetragonal distortion.
For $\vec{q}_{\Gamma \text{X}}(\xi)$ and $\vec{q}_{\Gamma \text{Z}}(\xi)$ 
the parameter $\xi$ varies between $0 \leq\xi\leq 1$ 
and 
for $\vec{q}_{\text{XP}}(\xi)$,$\vec{q}_{\text{XY}}(\xi)$, $\vec{q}_{\text{ZY}_{1}}(\xi)$
its range is  $0 \leq\xi\leq 0.5$.
Because of the higher symmetry of the fcc latice the
directions are reduced to 
$\vec{q}_{\Gamma \text{X}}(\xi)=\vec{q}_{\Gamma \text{Z}}(\xi)$ 
and 
$\vec{q}_{\text{XW}}(\xi)=\vec{q}_{\text{XP}}(\xi)=\vec{q}
_{\text{XY}}(\xi)=\vec{q}_{\text{ZY}_{1}}(\xi)$, accordingly.
In previous DFT
studies \cite{Knoepfle2000,Spisak2000,Spisak2002,Marsman2002,Sjostedt2002,Abrikosov2007}, 
SSs with $\xi=0.5,0.6$ 
for directions $\vec{q}_{\Gamma \text{X}}(\xi)$, $\vec{q}_{\Gamma \text{Z}}(\xi)$,
and $\xi=0.1,0.2$  for directions
$\vec{q}_{\text{XW}}(\xi)$ and the related directions
$\vec{q}_{\text{XP}}(\xi)$, $\vec{q} _{\text{XY}}(\xi)$, 
$\vec{q}_{\text{ZY}_{1}}(\xi)$ were found to be in contest.
At each point of the magnetic structure map the choice of propagations 
was made as just discussed.  At each of the three energy minima of the
map (see Fig.\ref{fig4} and Table \ref{tab1})
a much finer scan of $\vec{q}(\xi)$-vectors in steps of $\Delta \xi =0.01$ was made.
In addition, the accuracy of the generalized Bloch theorem  in
comparison to suitable supercell calculations  was tested and found to be
sufficient: the differences of total energies between both approaches
were always $\leq 0.3$  meV/atom. 

For all investigated spin structures Fig. \ref{fig3} depicts $\Delta E_\text{reor}$ 
depending on $c/a$. For each point and configuration the
energy was minimized with regards to $V$. The two regions $c/a<1$ and
$c/a>1$  are clearly distinguishable by the most stable spin structures.  For
$c/a>1$ SS configurations $\vec{q}_{\Gamma \text{Z}}$, $\vec{q}_{\text{Z}
\text{Y}_1}$ propagating along the $c$ axis and collinear structures AFM$_Z$
are favored. For $0.93<c/a<1$ clearly one structure is
most stable, namely the newly found AFM/NM ordering (see Fig.~\ref{fig1}). 

Discussing the volume dependency 
the collinear configurations AFM, AFM/NM,
L1$_2$ and the non-collinear SSs
$\vec{q}_\text{XP}(\xi)$,  $\vec{q}_\text{XY}(\xi)$, 
$\vec{q}_{\text{ZY}_1}(\xi)$ have their respective minima of $\Delta E_\text{reor}$
in the range of $10.4 \leq V \leq  10.8$ \AA$^3$. For
dl-AFM   and the SSs
$\vec{q}_{\Gamma\text{X}}(\xi)$, $\vec{q}_{\Gamma\text{Z}}(\xi)$  the minimum
of $\Delta E_\text{reor}$ appears at the larger volumes 
$10.6 \leq V \leq 11.3$ \AA$^3$.
A ferromagnetic low-moment (LM) phase with a moment of
$\mu=0.99\,\mu_\text{B}$ appears at the minimum with $V=10.5~$ \AA$^3$ and  $c/a=1$.  
For $0.95<c/a<1.075$ the LM 
ferromagnetic configuration is more favorable than the two  high-moment
(HM) ferromagnetic phases which are a) an fct phase with 
$\mu=2.35\,\mu_\text{B}$ and its minimum at  $V=11.7$ \AA$^3$, 
$ c/a=1.175$,
and b)  the HM bcc $\alpha$ phase with $\mu=2.16\,\mu_\text{B}$
at $c/a=1/\sqrt{2}$) and $V=11.3~$ \AA$^3$
 (see Table~\ref{tab1}).

The centerpiece of our work is shown in Fig.~\ref{fig4}, presenting the
structure map of magnetic phase stability as a function of volume and $c/a$
ratio. It combines the results for collinear orderings and SSs
in terms of the lowest $\Delta E_\text{reor}$.
Three local minima were found (see Table~\ref{tab1}) as marked by
\textbf{x1}, \textbf{x2}, and \textbf{x3} (see Fig.~\ref{fig4}).  
The minima \textbf{x1} and \textbf{x2} occur for $c/a>1$ 
whereas \textbf{x3} is found for $c/a<1$.  If only SSs are
considered the two local minima SS1 and SS2 appear as listed in
Table~\ref{tab1}.
SS1 with $\displaystyle \vec{q}_{\text{ZY}_1}(0.1)$ has
its  minimum for $c/a>1$ whereas for SS2 with 
$\displaystyle \vec{q}_\text{XP}(0.2)$ the minimum is 
for $c/a<1$.  

\begin{table}
\caption{
\label{tab1}
Magnetic orderings and corresponding volumes, $c/a$ ratios and
re-orientation energies $\Delta E_\text{reor}$. 
First three lines: the three local minima (see Fig.~\ref{fig4}). 
Fourth and fifth line: minima of most stable SSs.
Last four lines: minimized $\Delta E_\text{reor}$ for high-moment (HM) and
low-moment (LM) ferromagnetic ordering, the non-spinpolarized 
(NM) calculation, the bcc FM $\alpha$ phase, and the
L1$_2$ structure.
}
\begin{tabular}{ll|c|c|c}
&mag. ord. &    \quad$V$\quad  &\quad$c/a$\quad&  $\Delta E_\text{reor}$ \\
&          & [\AA$^3$]&      &  [meV/atom] \\
\hline
\hline
\textbf{x1} &AFM$_\text{Z}$   & 10.7 & 1.075 &-52\\
\textbf{x2} &dl-AFM$_\text{Z}$& 11.2 & 1.100 & -51\\
\textbf{x3} &AFM/NM           & 10.6 & 0.975 & -48\\
\hline
SS1   &$\displaystyle \vec{q}_{\text{ZY}_1}(0.1)$  & 10.7&1.075 &-51\\ 
SS2   &$\displaystyle \vec{q}_\text{XP}(0.2)$ & 10.7&0.950 &-42\\ 
   \hline
   &FM (LM) & 10.5 &1.000   & \quad0 \\
   &L1$_2$  & 10.7 &1.000& -42 \\   
   &FM (HM) & 11.7 &1.175 &-25  \\
   &NM     & 10.2 &1.000 &~19\\
   &FM bcc Fe&11.3&1/$\sqrt{2}$&-136
\end{tabular}
\end{table}

Minimum \textbf{x1} with $c/a=1.075, V=10.7$~\AA$^3$ represents the
collinear AFM$_\text{Z}$ configuration. However, Table~\ref{tab1}
(lines one and four) shows that the energy difference between AFM$_\text{Z}$
and SS1 with $\vec{q}_{\text{ZY}_1}(0.1)$ is only 1 meV. In fact,
a small orthorhombic distortion with $b/a\sim1.015$ stabilizes SS1 by 0.2
meV/atom  as predicted by Marsman {\em et al.}\cite{Marsman2002} and confirmed
experimentally by Tsunoda {\em et al.}.\cite{Tsunoda2007}  For $1<c/a<1.075$
SS1 is always  more favorable than AFM$_\text{Z}$ but for $c/a
\geq 1.075$ AFM$_\text{Z}$ is more favorable than any SS1 with
$\vec{q}_{\text{ZY}_1}(\xi)$ and $\xi>0$,
as stated in Ref.~\onlinecite{Marsman2002}.

Minimum \textbf{x2} with $c/a=1.10, V=11.2$ \AA$^3$ belongs to 
dl-AFM$_\text{Z}$. At these coordinates the closest competing
configuration is the SS with $\vec{q}_{\Gamma\text{Z}}(\xi=0.6)$ which is less
stable by 8 meV/atom.  This result was confirmed by calculating  SSs for $0<\xi<1$.
For dl-AFM$_\text{Z}$  no atomic relaxation were considered,
which would further lower $\Delta E_\text{reor}$.  Therefore,
in contrast to Refs.  \onlinecite{Abrikosov2007,Sjostedt2002} we exclude that
any SS will be more stable than dl-AFM$_\text{Z}$ 
at volumes larger than 11\AA$^3$.
The collinear configurations dl-AFM$_\text{X}$ and dl-AFM$_\text{Z}$ are
the dominating structures but they are unstable against
monoclinic shearing.~\cite{Spisak2002,Marsman2002}

Minimum \textbf{x3} corresponding to the AFM/NM configuration 
with its peculiar mixture of AFM and NM
planes (see Fig. \ref{fig1}) 
is the shallowest one (see Table~\ref{tab1}).  Nevertheless, it is the only
configuration with a local minimum for $c/a<1$, namely 
$c/a=0.975, V=10.6$~\AA$^3$.
Supposedly, the AFM/NM
configuration indicates formation of an SS.  However, the corresponding
SSs with propagations $\vec{q}_{\Gamma\text{X}}(0.5)$ and
$\vec{q}_{\Gamma\text{Z}}(0.5)$ are very unfavorable
for this particular $c/a$ (see Fig. \ref{fig3}):  
AFM/NM is by 9 meV/atom more stable than the closest
non-collinear ordering SS2 with $\vec{q}_\text{XP}(\xi=0.2)$. 
Varying $\xi$
at the same $c/a$ and $V$ shows that indeed 
SS2 with $\xi=0.2$ is the most favorable SS. 
Presumably the AFM/NM configuration has been
detected previously by LEED measurements at 300K on thin
films consisting of  10 to 12 mono-layers.  \cite{Lu1989,Darici1987}
Subsequent LEED experiments \cite{Landskron1991,Wutting1993}
observed a distinct orthorhombic distortion and volume expansion \cite{Wutting1993} 
when the samples were further cooled down
resembling a transition from minimum \textbf{x3} to \textbf{x1}.
The analysis of the experimental results was rather inconclusive
with respect to the magnetic ordering, and a range of magnetic
configurations from nonmagnetic to ferromagnetic to anti-ferromagnetic
orderings were suggested.
\cite{Pescia1987,Liu1988,Macedo1988,Stampanoni1989,Wutting1993} 
These observations, while seemingly contradicting each other support our
finding of the AFM/NM configuration. 

The stability of AFM/NM compared to L1$_2$ is
illustrated by the density of states (DOS) (see Fig.~\ref{fig1}): 
the values of the DOS  at
Fermi energy, $\mathrm{N(E_F})$  for both spin channels of L1$_2$ 
is larger by 40\% than for AFM/NM (see also Supplementary B).
For L1$_2$  the spin up and  down DOS is not
symmetric and the total moment is not zero. This is in contrast to 
AFM/NM  for which the total moment is zero because for each
layer perpendicular to the $c$-axis the local moments 
$\mu=\pm 1.8 \,\mu_\text{B}$) 
either cancel  or are perfectly zero. 
Performing studies with different spin splits (see Supplementary B)
it turns out that the stability of AFM/NM is due to its
lowest $\mathrm{N(E_F})$. By orthorhombic distortion a structure is stabilized
for which the magnetically dead atoms accumulate finite local
moments. Its crystal structure resembles the structure at minimum \textbf{x1} 
(see Supplementary A).

\begin{figure}
\includegraphics[width = 0.4\textwidth ]{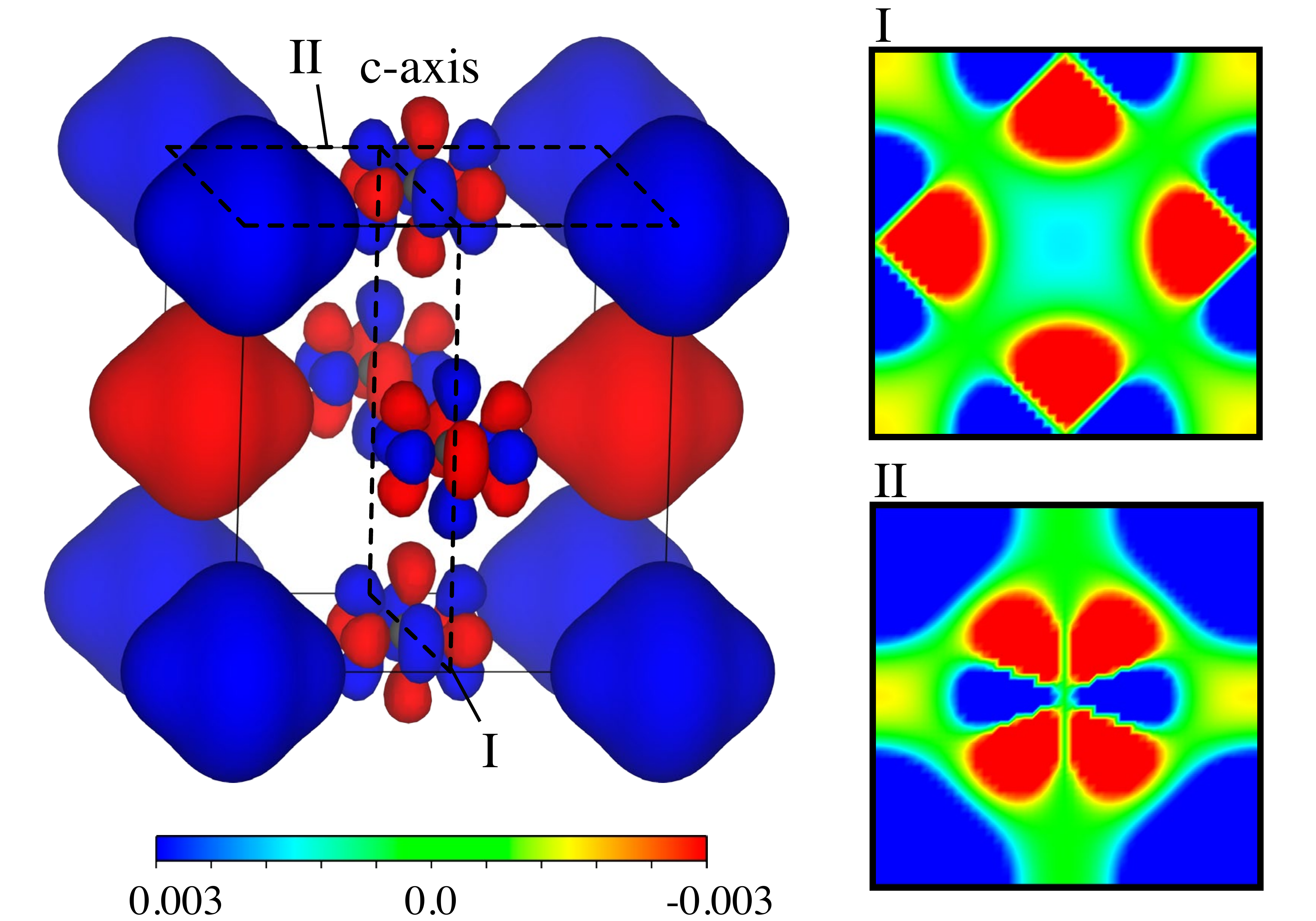} 
\caption{(Color online) 
Magnetization density $\rho_\text{mag}=\rho_\text{up}-\rho_\text{down}$
($\rho_\text{up},\rho_\text{down}$: spin up and down charge densities) of
AFM/NM 
ordering.
Left panel: three-dimensional mantle;
right
panel: contour plots in planes
as sketched  in left panel. Figure created by VESTA\cite{Momma2011}.
}
\label{fig5} 
\end{figure}

The peculiarity of AFM/NM is illustrated by
Fig.\ref{fig5} showing that the magnetization density around the positions of
the magnetically dead atoms is strongly spin polarized in a symmetric manner
such that the resulting local moments are zero.
This symmetry property remains even when
the AFM/NM structure is tetragonally distorted according to Fig.~\ref{fig1}. 
Consequently, AFM/NM 
is
the most stable spin ordering for $0.94 \le c/a \le 1.01$. 

Summarizing, our extensive search for magnetic configurations of fct Fe in
terms of a magnetic structure map predicts a range of magnetic orderings.  In
particular, on this energy landscape depending on volume per atom and 
$c/a$ ratio a hitherto unknown and simple collinear
anti-ferromagnetic ordering with magnetically dead Fe atoms was found.  We
believe by that the riddle concerning magnetic ordering and structure as posed
by experiment is 
finally solved.

\acknowledgments
Work was supported by the Austrian Science Fund FWF within the Special
Research Program VICOM (Vienna Computational Materials Laboratory, Project No.
F4110). Calculations were done on the Vienna Scientific Cluster (VSC).

\end{document}